\documentclass[journal=jacsat,manuscript=article]{achemso}

\usepackage[version=3]{mhchem} 

\usepackage{xcolor}
\graphicspath{ {./images/} }

\DeclareUnicodeCharacter{2212}{-}
\DeclareUnicodeCharacter{0301}{\'{y}}

\title[An \textsf{achemso} demo]
  {Investigation of the Nonlinear Optical Frequency Conversion in Ultrathin Franckeite Heterostructures}

\author{Alisson R. Cadore}
\affiliation{Mackenzie Engineering School, Mackenzie Presbyterian University, Sao Paulo, 01302907, SP, Brazil.}
\alsoaffiliation {Brazilian Nanotechnology National Laboratory, Brazilian Center for Research in Energy and Materials, Campinas, 13083-100, SP, Brazil.}
\alsoaffiliation {These authors contributed equally to this work.}

\author{Alexandre S. M. V. Ore}
\affiliation{Mackenzie Engineering School, Mackenzie Presbyterian University, Sao Paulo, 01302907, SP, Brazil.}
\alsoaffiliation {MackGraphe, Mackenzie Presbyterian Institute, Sao Paulo, 01302907, SP, Brazil.}
\alsoaffiliation {These authors contributed equally to this work.}

\author{David Steinberg}
\affiliation {MackGraphe, Mackenzie Presbyterian Institute, Sao Paulo, 01302907, SP, Brazil.}

\author{Juan D. Zapata}
\affiliation{Faculty of Engineering, Universidad del Antioquia UdeA, Medell\'{i}n, Colombia}

\author{Eunézio A. T. de Souza}
\affiliation{Mackenzie Engineering School, Mackenzie Presbyterian University, Sao Paulo, 01302907, SP, Brazil.}

\author{Dario A. Bahamon}
\affiliation{Mackenzie Engineering School, Mackenzie Presbyterian University, Sao Paulo, 01302907, SP, Brazil.}
\alsoaffiliation {MackGraphe, Mackenzie Presbyterian Institute, Sao Paulo, 01302907, SP, Brazil.}

\author{Christiano J. S. de Matos}
\affiliation{Mackenzie Engineering School, Mackenzie Presbyterian University, Sao Paulo, 01302907, SP, Brazil.}
\alsoaffiliation {MackGraphe, Mackenzie Presbyterian Institute, Sao Paulo, 01302907, SP, Brazil.}
\email{cjsdematos@mackenzie.br}

\begin{document}

\begin{abstract}
  Layered franckeite is a natural superlattice composed of two alternating layers of different compositions, SnS$_2$- and PbS-like. This creates incommensurability between the two species along the planes of the layers, resulting in spontaneous symmetry-break periodic ripples in the \textit{a}-axis orientation. Nevertheless, natural franckeite heterostructure has shown potential for optoelectronic applications mostly because it is a semiconductor with 0.7 eV bandgap, air-stable, and can be easily exfoliated down to ultrathin thicknesses. Here, we demonstrate that few-layer franckeite shows a highly anisotropic nonlinear optical response due to its lattice structure, which allow for the identification of the ripple axis. Moreover, we find that the highly anisotropic third-harmonic emission strongly varies with material thickness. These features are further corroborated by a theoretical nonlinear susceptibility model and the nonlinear transfer matrix method. Overall, our findings help to understand this material and propose a characterization method that could be used in other layered materials and heterostructures to assign their characteristic axes.
\end{abstract}

\section{Introduction}

Layered materials (LMs) have shown interesting electrical, optical, and mechanical properties\cite{choi2017recent,backes2020production,Barcelos2020,deMatos2023} and are of great technological importance\cite{akinwande2014two,liao2019van}. Many of these LMs are naturally occurring\cite{riccardo2020naturally,santos2019exfoliation,deOliveira2022,cadore2022exploring,JOSA2023,Longuinhostalc,Review23}, while others can be synthesized\cite{mannix2017synthesis,shivayogimath2019universal,HamzaJAP22,Nagaoka23}. Two or more LMs can, nowadays, be assembled together by deterministic stacking or directly grown by chemical vapour deposition (CVD) and epitaxial growth techniques, creating the so-called vertical van der Waals heterostructures (vdWHs)\cite{geim2013van,liu2016van}. Such vdWHs offer a platform to obtain unprecedented functionalities and properties that do not exist in the individual LMs\cite{Andrei2021,Novoselov_vdWHs}. Albeit these vdWHs are well-studied and applied\cite{mak2016photonics,sun2016optical,Feres2023,Nutting2021,liu2016van,Gadelha_2019}, the fabrication process to create these nanostructures suffers from some drawbacks such as the need for expensive systems for the growth, or the presence of contaminants between the layers, besides the difficulty of depositing the LMs with high-precision, in the manual stacking case\cite{Novoselov_vdWHs,chakraborty2022challenges,montblanch2021confinement,geim2013van,roman2020tunneling}. To solve these issues, mother nature has given us many minerals that are naturally occurring vdWHs. These vdWHs are formed with alternating layers of two different LM due to the phase segregation process during their formation, and thus are free of the aforementioned problems. So far, several exfoliated and naturally occurring vdWHs have been identified, such as cylindrite\cite{Niu_2019,dasgupta2021natural,molecules26237371}, lengenbachite\cite{dasgupta2021naturally}, levyclaudite\cite{chen2019natural}, nagayagile\cite{dasgupta2021naturally}, teallite\cite{gusmao2018layered,SHU2020_teallite}, to name a few, and the most notably studied, franckeite\cite{williams1988electron,costa2021vibrational,Dunieskys_2021,Zschiesche_2021,paz2021franckeite,molina2017franckeite,velicky2017exfoliation,gant2017optical,frisenda2020symmetry,XU2023}.

Franckeite is a structure composed of alternating stacks of PbS-type pseudotetragonal (Q) layers and SnS$_2$-type  pseudohexagonal (H) layers, as shown in Fig. \ref{fig:flake}a\cite{williams1988electron,wang1991crystal,henriksen2002atomic}. So far, it has been shown that the alignment of each franckeite monolayer (Q-H layers) is preserved in the exfoliated flakes (i.e. Q-H/Q-H/.../Q-H)\cite{molina2017franckeite,velicky2017exfoliation}. Nevertheless, due to its natural occurrence, the material may have extra impurities in the crystal lattice which can cause Q and H layer stacking faults (i.e. Q-H/H/Q-H or Q-H/Q/Q-H...) which are random in the material, consequently, modifying the electronic properties of the material on a specific flake location\cite{Zschiesche_2021}. However, to the best of our knowledge, the direct correlation between stacking faults and material properties was not assigned, and this could explain some data deviations between research groups working on naturally obtained LMs. Franckeite is also a p-type semiconductor with a bandgap of approximately 0.7 eV\cite{frisenda2020symmetry, molina2017franckeite,velicky2017exfoliation} and it has gained notoriety for its potential for optoelectronic applications\cite{Ray_2017_FranckPhoto,Li2020_passiveFranck,molina2017franckeite,frisenda2020symmetry,Costa_PL2022}. Other works have been carried out to understand its structural and optical properties. For instance, studies have characterized its vibrational properties \cite{costa2021vibrational,frisenda2020symmetry}, its optical absorption\cite{XU2023}, and its refractive index, by means of optical contrast\cite{gant2017optical}, as well as its third-order nonlinear optical properties, characterized by Z-scan\cite{Li2020_passiveFranck,Jie_Franck_LPE} and third harmonic generation (THG)\cite{tripathi2021naturally} techniques. In particular, Tripathi and co-workers\cite{tripathi2021naturally} have reported on the polarization and thickness dependence of THG. However, few-layer thicknesses were not reached and, thus, they did not notice any polarization-dependence changes with thickness.  

It is also known that franckeite exhibits spontaneous symmetry break due to the incommensurability of its lattice network\cite{velicky2017exfoliation,molina2017franckeite,williams1988electron,wang1991crystal,henriksen2002atomic}. This process introduces a strain-induced structural deformation, which is known to develop symmetry-break periodic ripples in the $a$ axis orientation. The rippling arises from the competition between inhomogeneous van der Waals interlayer interactions and inhomogeneous elastic deformations\cite{frisenda2020symmetry,Zschiesche_2021}. As the crystal relaxes mechanically to minimize the total energy, it develops both strong in-plane strains and a small out-of-plane rippling, modulated along the armchair direction of the H layer \cite{frisenda2020symmetry,Zschiesche_2021}. These ripples, as well as the $a-b$ plane of franckeite have been deeply investigated by means of electron microscope techniques\cite{paz2021franckeite,frisenda2020symmetry,velicky2017exfoliation,tripathi2021naturally,molina2017franckeite,williams1988electron,wang1991crystal,henriksen2002atomic,paz2021franckeite} and, more recently, by polarization-dependent Raman spectroscopy\cite{frisenda2020symmetry,tripathi2021naturally}. However, these techniques commonly require long operational times. Therefore, it is still desired to demonstrate a tool that is non-destructive, and that is able to quickly assign the ripples´ axis. Thus, in this work, we demonstrate a fast and simple way to assign the franckeite ripples' axis. It consists of determining the polarization-dependent THG in the natural heterostructure. Thus, it is similar, requiring an alike experimental setup, to that widely used for determining the crystal axes of 2H-phase transition metal dichalcogenides (in the latter case, however, second-harmonic generation (SHG) needs to be used)\cite{vianna2021second,Sousa_2021}. The method also determines the anisotropic response of franckeite's third-order nonlinear susceptibility. We show that not only the rippling direction can be identified, but also that the polarization-dependent THG response of franckeite shows noticeable dependence on the sample thickness, which is well explained by our theoretical nonlinear susceptibility model. Most importantly, we determine  the elements of third-order susceptibility tensor ($\chi^{(3)}$) and find that they undergo noticeable changes for very thin flakes.
We envisage that this experimental procedure and results will contribute to the understanding of naturally occurring vdWHs and could potentialize the use of these low-cost LMs in future photonic and optoelectronic applications.

\section{Methods}
\subsection{Sample Preparation}

A bulk-franckeite sample from San Jose mine, Oruro - Bolivia, was mechanically cleaved by scotch tape and then exfoliated again on a polydimethylsiloxane (PDMS) stamp placed on a glass slide for inspection under an optical microscope\cite{castellanos2014deterministic}. Optical contrast was optimized to identify few-layer franckeite flakes prior to transfer\cite{gant2017optical}. The selected samples were then stamped on a 500 $\mu$m glass thick substrate at 40°C using \textit{xyz} micro-manipulators under an optical microscope, before increasing the temperature to 60°C, so the flakes detach from the PDMS and adhere preferably to the glass surface.

\subsection{Sample Characterization}

After the flake transfer, Atomic Force Microscopy (AFM - Bruker Dimension Icon system) was performed in tapping mode to characterize the sample height. Franckeite flakes with thicknesses ranging from 8 nm to 130 nm thick were transferred and analyzed. These thicknesses were identified by AFM measurements in two staircase-like franckeite samples that were investigated in this work. For the Raman measurements, a confocal Raman microscope (WITec Alpha 300R system) was used in the backscattering configuration at 532 nm excitation energy with a 100X objective which has a numerical aperture NA = 0.9. Each Raman spectrum was collected with three accumulations of 30 seconds with a grating of 1800 grooves/mm. The laser power was kept constant at $\sim$1 mW during all measurements and the laser was previously aligned on a prime silicon wafer to use the Si peak at $\sim$521 cm$^{-1}$ as reference\cite{ardito2018damping}. Polarized Raman measurements were also performed on selected flakes. For this, an optical linear polarizer (analyzer) was added to the reflected (collection) path of the microscope before the signal reaches the spectrometer\cite{marangoni2021long}. We fixed the laser's incident polarization as parallel to the analyzer axis. To collect the polarized Raman spectrum as a function of the angle, we stepwise rotated the sample stage using a goniometer, acquiring one spectrum per step.

\begin{figure*}[ht]
 \centering
 \includegraphics[width=\textwidth]{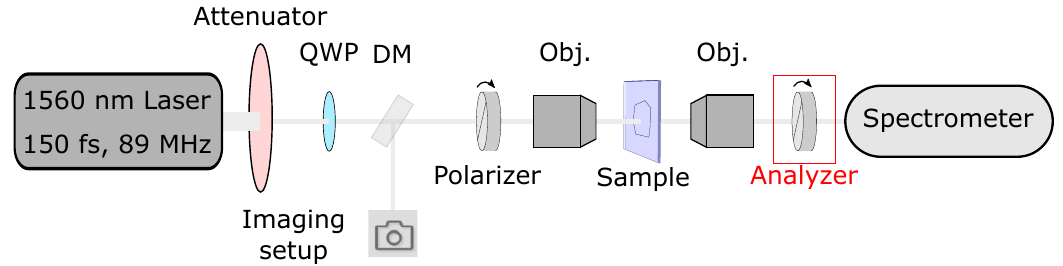}
 \caption{Experimental setup for THG characterization of the samples in transmission mode. QWP: quarter-wave plate; DM: dichroic mirror; Obj.: microscope objective. The red box indicates that the analyzer can be used or removed from the setup, depending on the desired experiment.}
 \label{fig:setup}
\end{figure*}

The THG experimental setup is illustrated in Fig. \ref{fig:setup}. It is composed of a mode-locked Erbium-doped fiber laser (at a wavelength of 1560 nm, yielding 150-fs pulses at a repetition rate of 89 MHz) that is focused on the sample at normal incidence through a 20X objective lens (NA = 0.5). The sample is placed on an $xyz$ translation stage that is controlled by piezo-scanners. The average laser power used in the experiments and measured before the objective is 10 mW, except in the THG power dependence measurements. The $1/e^2$ beam diameter was measured to be 3.6$\mu$m\cite{Woodward_2016}. The THG signal at 520 nm is collected in transmission mode, passing through another objective lens, which collimates the beam, and directs it to a spectrometer (Andor Kymera coupled to a silicon CCD iDus 416 camera)\cite{vianna2021second}. A variable optical attenuator, at the laser output, controls the laser power while a quarter-wave plate (QWP) followed by a linear polarizer in a motorized rotation stage are used to ensure a practically constant pump laser intensity (with power variation below 2\%) independently of the polarization angle. The franckeite flakes, as well as the THG spot, were imaged in reflection by a microscope setup, consisting of a white LED, a CCD camera, and some simple optics (Fig. \ref{fig:setup}) so that the laser position in the sample could be identified (Fig. \ref{fig:flake}b-d).

For polarization-resolved measurements, the motorized stage containing the polarizer was rotated in $5^{\circ}$ steps from $0^{\circ}$ to $360^{\circ}$.  
Two types of polarization-dependent THG measurements were performed: i) an analyzer placed after the second objective was inserted and remained at a (chosen) fixed angle, while the polarizer was rotated;  
ii) no analyzer was used while the polarizer was angle scanned. 

To obtain the THG signal as a function of pump power, we selected a representative franckeite region, with homogeneous thickness. Before collecting the spectra, we maximized the THG signal by changing the sample focus on the flake surface. Then, the pump power was tuned from 1 mW to 11 mW and each spectrum was taken with two accumulations of 1 second acquisition time. In this case, the polarizer and analyzer were kept parallel to each other, at a fixed angle. The results presented for each pump power were averaged from five spectra. 

To acquire the THG signal as a function of thickness, we carried out the following steps, for each region: i) the THG signal was maximized by adjusting the sample-objective distance; ii) full polarization-dependent THG measurements were conducted; iii) we selected the angle with maximum THG intensity and averaged the THG signal over five spectra. All THG points presented in the manuscript, even those that were taken from different samples, were collected using these steps. Finally, note that the pump photon energy is beyond franckeite bandgap energy, leading to linear absorption, in addition to THG. Despite the fact that pump absorption was found to somewhat reduce transmittance in the characterized flakes ($\sim$90\% and $\sim$65\% transmittance for the thicker and thinner flakes measured, respectively), no detectable damage was observed throughout the experiments.

\subsection{THG power as function of thickness: theoretical approach}
 
To theoretically investigate the THG power as function of the thickness of the sample, we use the transfer matrix method for harmonic generation\cite{bethune1989optical}. Within this approach, the sample is modeled as a  nonlinear slab of thickness $d$  between two linear half-spaces. The THG fields that radiate to air, thus counter-propagating with the pump (region 1), and to glass, co-propagating with the pump (region 3), are then easily obtained as:

 \begin{eqnarray}
\begin{pmatrix}
E^+_3  \\
E^-_1
\end{pmatrix}=\frac{1}{1+\phi^2_2r_{12}r_{23}}
\begin{pmatrix}
\phi_2t_{23}[S^+ + r_{12}S^-]\\
t_{21}[\phi^2_2 r_{23}S^+ - S^-]
\end{pmatrix},
\label{eq:THGE3}
\end{eqnarray} 

\noindent where the refractive index $N_i$ of slab $i$, the terms $r_{ij}=\frac{N_i-N_j}{N_i+N_j}$, $t_{ij}=\frac{2N_i}{N_i+N_j}$ ($j=1,2,3$) and $\phi_2=e^{iN_2(3\omega/c)d_2}$ are evaluated at frequency $3\omega$. The source terms 

 \begin{eqnarray}
\begin{pmatrix}
S^+  \\
S^-
\end{pmatrix}=\frac{1}{t_{2s}}
\begin{pmatrix}
\phi^{-1}_2\phi_s - 1 & [\phi^{-1}_2\phi^{-1}_s - 1]r_s \\
[\phi_2\phi_s - 1]r_s& \phi_2\phi^{-1}_s - 1 
\end{pmatrix}
\vec{E_{2s}}
\label{eq:Es}
\end{eqnarray}

\noindent are related to the pump field in the nonlinear slab (layer 2) by: 

 \begin{eqnarray}
 \vec{E_{2s}} = \frac{1}{t^{\omega}_{21}}
\begin{pmatrix}
1+r^{\omega}_{21}r^{\omega} \\
r^{\omega}_{21}+r^{\omega}
\end{pmatrix}\left[\frac{4\pi}{\varepsilon_s}\right]\vec{P}_{NL},
\label{eq:EmasEmn}
\end{eqnarray}
 
 \noindent where the superscript ${\omega}$ means that the quantity is evaluated at the pump frequency. $r^{\omega}$ is the overall reflection coefficient and can be calculated from the well-known Airy's formula for three media $r^{\omega}_{123}$ or by the recursion formula for multiple reflections. $\vec{P}_{NL}=\chi^{(3)]}.\vec{E^\omega}.\vec{E^\omega}.\vec{E^\omega}$ is the nonlinear polarization. Note that when the tensorial product in the nonlinear polarization is expanded, four types of waves appear with the $3\omega$ frequency: $e^{\pm i3N_2k_0z}$ and $e^{\pm iN_2k_0z}$. Each one of these waves can be related to components of the non-linear polarization source term with different $\vec{k}$ vectors and effective dielectric constants $\varepsilon_s = \varepsilon_2(3\omega)-\varepsilon_2(\omega)$ for the waves $e^{\pm i3N_2k_0z}$ and  $\varepsilon_s = \varepsilon_2(3\omega)-\varepsilon_2(\omega)/9$ for the terms $e^{\pm iN_2k_0z}$\cite{bethune1989optical}. We treat independently each one of these contributions  through the  subscript $s$ in equation \ref{eq:EmasEmn}; to obtain  the total electric field in region 3 all contributions are summed up. 

\section{Results and Discussion}

\begin{figure*}[ht]
 \centering
 \includegraphics[width=0.8\textwidth]{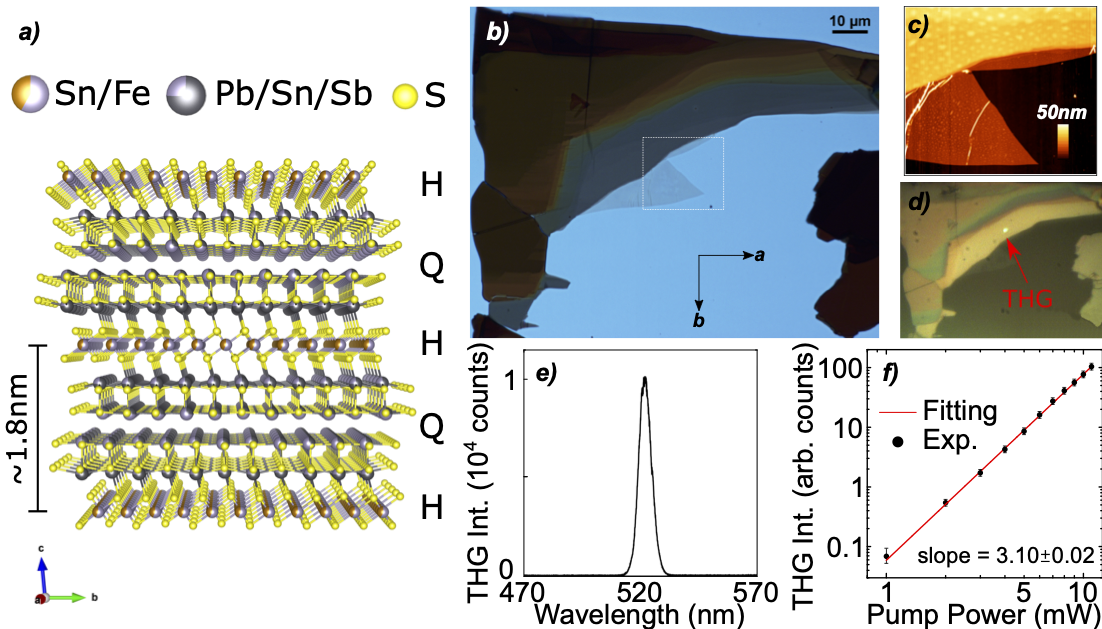}
 \caption{a) Crystal structure of franckeite where the two different stacked layers, the SnS$_2$-type (H) and the PbS-type (Q), can be seen.b) Transmission optical microscope image of an exfoliated staircase-like franckeite flake. Flake orientation ($a-b$) is indicated, in addition to the area where the few-layer franckeite was measured by AFM (white rectangle). c) AFM topography image collected from the area marked in (b). d) Lower resolution/contrast reflection image showing the THG emission (bright green spot) obtained with laser incidence onto a representative region. e) Measured THG spectrum with a peak wavelength at 520 nm from a 34 nm-thick franckeite flake. f) Double log-scale plot of the measured THG intensity as a function of the laser power with a power law fit ($\alpha=3.10\pm0.02$) to the data obtained from the region shown in (b).The error bars shown were calculated as the standard deviation of 5 measurements taken at the same laser power.}
 \label{fig:flake}
\end{figure*}

Fig. \ref{fig:flake}b) shows a transmission optical microscope image of one of the samples studied in this work, with the franckeite orientation ($a$ and $b$ axes) indicated on it. Regions of various colors are visible, which are caused by thin-film optical interference, indicating different flake thicknesses \cite{gant2017optical}, as confirmed by AFM measurements (Fig. \ref{fig:flake}c). Fig. \ref{fig:flake}d) shows the THG emission (green spot indicated by an arrow) captured by the CCD camera of the experimental setup, from a 34 nm-thick franckeite region during the measurements. A representative THG spectrum is plotted in Fig. \ref{fig:flake}e), demonstrating the peak wavelength at 520 nm which is exactly one-third of the excitation wavelength. Further confirmation of THG can be seen in Fig. \ref{fig:flake}f), which shows the pump power dependence of the THG intensity, with a cubic power law fit ($\alpha=3.10\pm0.02$) to the experimental data.

\begin{figure}[ht]
\centering
  \includegraphics[width=\textwidth]{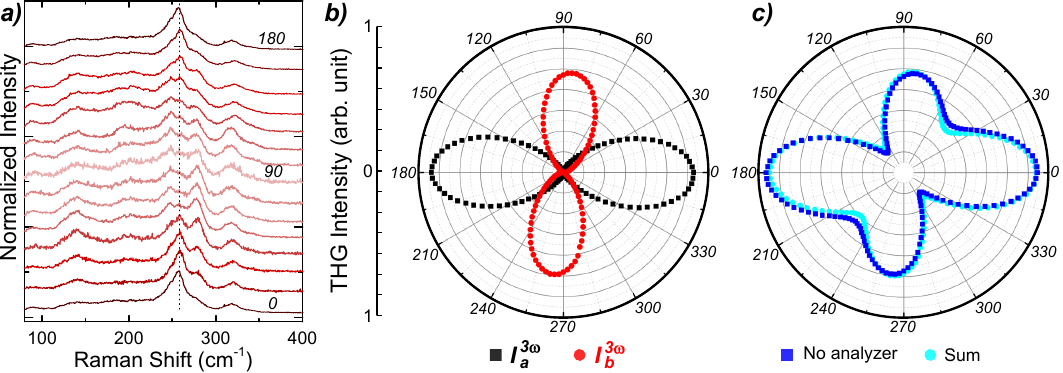}
  \caption{a) Representative angle-resolved polarized Raman spectra for a 34 nm-thick franckeite flake collected in the parallel polarization configuration under 532 nm laser excitation. The dashed black line indicates the most sensitive Raman mode to the laser polarization. b-c) THG intensity dependence on the incident linear polarization angle for a representative franckeite flake with 34 nm thickness, collected with (b) and without (c) the analyzer. 0$^\circ$ corresponds to the \textit{a}-axis (ripples direction). The red ($I{_a}^{3\omega}$) and black ($I{_b}^{3\omega}$) symbols in (b) correspond to the measurements carried out with the analyzer aligned parallel and perpendicular to the rippling axis, respectively. Part (c) also shows the trace obtained by the sum of the curves in (b) multiplied by 0.25 (i.e.,  $(I{_a}^{3\omega} +I{_b}^{3\omega} )\times0.25$), which accounts for the 25\% attenuation of the analyzer, as stated in its specifications.}
  \label{Fig:dep_pol}
\end{figure}

As mentioned before, franckeite is a natural vdWH with lattice incommensurability between the two consecutively stacked PbS-like and SnS$_2$-like layers\cite{velicky2017exfoliation,molina2017franckeite,williams1988electron,wang1991crystal,henriksen2002atomic}. This arrangement introduces a strain-induced structural deformation, creating the known periodic ripples in the $a$ axis orientation. Considering that this feature can be identified through polarized Raman spectroscopy\cite{frisenda2020symmetry,tripathi2021naturally}, we show in Fig. \ref{Fig:dep_pol}a a set of Raman spectra at various polarization angles collected from a 34 nm-thick flake. A typical Raman spectrum of franckeite exhibits several distinct Raman modes within the 75–400 cm$^{−1}$ window, which varies differently with polarization\cite{frisenda2020symmetry,tripathi2021naturally}. These previous works demonstrated that the peak at $\sim$265 cm$^{−1}$ (highlighted by the vertical dashed line in Fig. \ref{Fig:dep_pol}a) is the most polarization-sensitive mode and can be used to assign the rippling orientation of franckeite. In particular, the maximum (minimum) mode intensity is observed when the laser polarization is parallel (orthogonal) to the rippling axis ($a$ axis). Hence, the 0$^\circ$ (90$^\circ$) polarization angle is defined as parallel (perpendicular) to the $a$ axis (ripples). Note that we have used this approach to identify the franckeite orientation, as shown in Fig. \ref{fig:flake}b, prior to further experiments. It is important to stress that polarized Raman measurements taken on different franckeite thicknesses (Fig. \ref{fig:flake}c) of the same flake exhibit a similar trend for the Raman modes as a function of the polarization angle, suggesting that if a flake is composed of regions with different layer numbers (as the one in Fig. \ref{fig:flake}b-d), the orientation of the ripples along the entire flake is maintained. Therefore, by finding the rippling orientation in one region of the flake, one can assume the same alignment for all regions.

\begin{figure}[ht]
\centering
  \includegraphics[width=0.8\textwidth]{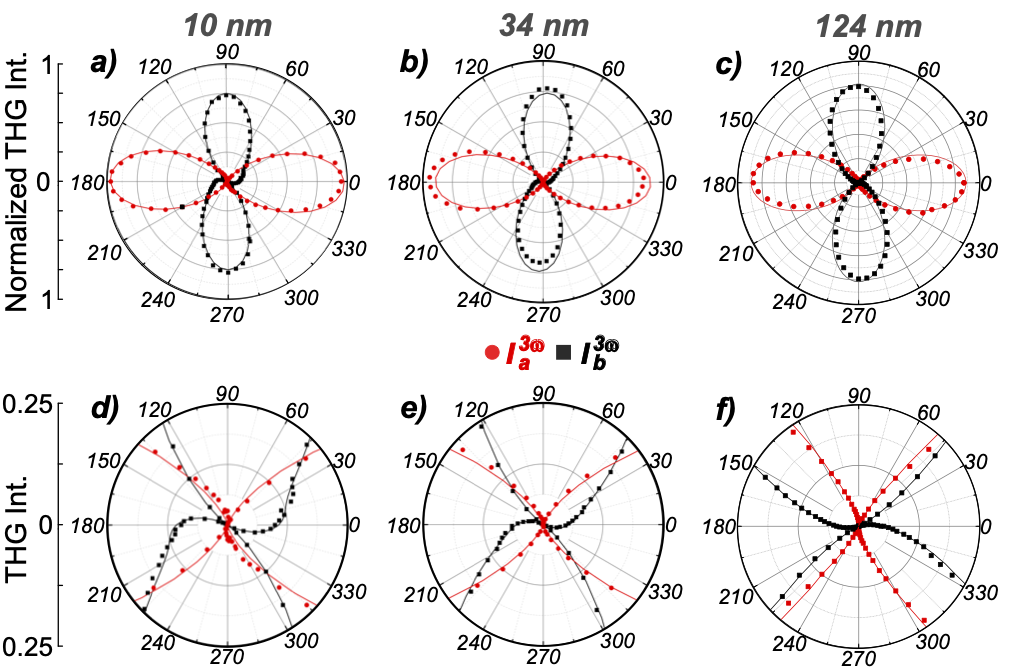}
  \caption{Angular dependence of the THG intensity on the incident linear polarization angle for franckeite flakes with thicknesses of (a) 10 nm, (b) 34 nm, and (c) 124 nm. d-f) are zoomed graphs close to the lobe bases of the respective samples. Red and black symbols correspond to the measurements carried out with the analyzer aligned parallel and perpendicular to the rippling axis, respectively. The corresponding theoretical fittings are plotted as solid curves.}
  \label{Fig:few_mid}
\end{figure}

Since the linear optical response of franckeite shows strong in-plane structural anisotropy with the incident light’s polarization\cite{frisenda2020symmetry}, it can be expected that the nonlinear response would show similar behavior. We have then investigated the effect of this in-plane anisotropy of franckeite through polarization-dependent THG (see Methods for more details). We start the polarization-dependent THG characterization by fixing the analyzer (Fig. \ref{fig:setup}) either at the 0$^\circ$ or the 90$^\circ$ angle, as previously identified by Raman spectroscopy (Fig. \ref{Fig:dep_pol}a). Fig. \ref{Fig:dep_pol}b displays the angular dependence of the THG intensity on the incident laser polarization angle relative for a 34 nm-thick franckeite flake. With the analyzer parallel to the $a$ axis, a symmetric two-lobe pattern is observed (black symbols), with the maximum THG occurring at 0$^\circ$ (parallel to the rippling direction). When the analyzer is orthogonal to the $a$ axis, a second two-lobed curve (red symbols) is observed. As discussed later, this curve exhibits asymmetry at low THG intensity angles (see Fig. \ref{Fig:few_mid}a). The emission pattern shows the maximum THG occurring at 0$^\circ$ (parallel to the rippling direction) and the secondary maximum occurring close to 90$^\circ$.

The anisotropy ratio of the THG signal I${_a}^{3\omega}$($\theta=0^\circ$)/I${_b}^{3\omega}$($\theta=90^\circ$) is around 1.25. 
Note that the overall maximum THG is conveniently obtained along the rippling axis (with the analyzer along the $a$ axis), which means it can potentially be used for franckeite orientation determination. However, in the set of data shown so far, this axis orientation needs to be known \textit{a priori}, so that the analyzer was positioned. Nevertheless, we show in Fig. \ref{Fig:dep_pol}c (dark blue) that by performing a single polarization-dependent THG characterization without the analyzer, one gets directly the same lobe structure and, consequently, the rippling orientation. Indeed, the obtained angular profile matches that obtained by the sum of the THG profiles measured with the analyzer along the $a$ and $b$ axes, also shown in Fig. \ref{Fig:dep_pol}c (cyan). Additionally, the anisotropy ratio of the THG signal I${_a}^{3\omega}$($\theta=0^\circ$)/I${_b}^{3\omega}$($\theta=90^\circ$) $\sim$1.26 is recovered as well. It is worth mentioning that we have previously shown that the THG signal of franckeite can be $\sim$3 orders of magnitude higher than graphene \cite{Steinberg_Franck2018}; hence, one can take $<$1 min for identifying the franckeite axes without further characterization. Therefore, we demonstrate that this approach can be used to quickly identify the franckeite orientation. Additionally, it may potentially be applied to other anisotropic LMs and heterostructures\cite{tripathi2021naturally,dasgupta2021natural,sar2020plane,dasgupta2021naturally,dasgupta2020anisotropic,youngblood2017layer}.

We now focus on the THG dependence on the franckeite thickness. Figs. \ref{Fig:few_mid}a-c show representative THG profiles as a function of polarization, for three flake thicknesses: 10 nm, 34 nm, and 124 nm, respectively. In the polarization-dependent THG measurements, we observe differences in the lobe shape for angles close to the THG minima (Figs. \ref{Fig:few_mid}d-f). Note that, when the analyzer is parallel to the $b$ axis, the thinner the flake, the more pronounced is the shoulder-like feature close to the base of the lobes (lower THG intensities); for thicker samples, a more symmetric lobed structure is obtained. These changes with flake thickness are reproducible and beyond experimental errors, and seem to indicate a change in the $\chi^{(3)}$ tensor, as franckeite is thinned down. To quantify the $\chi^{(3)}$ variation, we use a nonlinear susceptibility model\cite{tripathi2021naturally}, and perform the fitting of these three representative experimental THG traces (solid lines in Fig. \ref{Fig:few_mid}a-f). Even though this procedure is common practice in the literature, we stress that accurate determination of the nonlinear tensor components is tricky and cannot be done univocally solely from THG measurements. Thus, the purpose here is just to provide fitting to the experimental curves, as well as to indicate which tensor components might dominantly change with thickness.

In the triclinic crystal system all elements of the third-order susceptibility  are independent and non-zero. Despite this, the tensor can be contracted  for a simplified computation of  the THG fields.  \cite{Yang:95}  The element $\chi^{(3)}_{ijkl}$ of the  fourth-rank tensor is expressed  in the contracted notation as $\chi^{(3)}_{mn}$. Here, the index $m=1,2,3$ denotes the spatial component along the directions $\hat{a},~\hat{b}~\text{or}~\hat{c}$ of the THG field. The contracted index  $n$ represents  permutations of the three spatial components of the incident  wave ($\hat{e}_x, \hat{e}_y~\text{and}~ \hat{e}_z $) and varies from 1 to 10. \cite{Yang:95} 
With these considerations, the THG field is

\begin{equation}
\left(\begin{matrix} E_a^{3\omega}\\E_b^{3\omega}\\E_c^{3\omega}\end{matrix}\right) \propto
\begin{pmatrix}
\chi_{11} & \chi_{12} & \chi_{13} & \chi_{14} & \chi_{15} & \chi_{16} & \chi_{17} &\chi_{18} &\chi_{19} & \chi_{110} \\
\chi_{21} & \chi_{22} & \chi_{23} & \chi_{24} & \chi_{25} & \chi_{26} & \chi_{27} &\chi_{28} &\chi_{29} & \chi_{210}  \\
\chi_{31} & \chi_{32} & \chi_{33} & \chi_{34} & \chi_{35} & \chi_{36} & \chi_{37} &\chi_{38} &\chi_{39} & \chi_{310} 
\end{pmatrix}
\cdot
\vec{L}.
\label{eq:MChi}
\end{equation}

\noindent  Where the rows of the column vector $\vec{L}$ are  the combinations of the three components of the pump field,  as shown below. For the analysis, we take into account that the pump laser has a fundamental frequency of 1560 nm and that the beam is linearly polarized. Moreover, as the experiment is carried out at normal incidence, the pump laser electric field at the focal point of the objective only presents components in  the $\hat{e}_x-\hat{e}_y$ plane. Consequently, we can disregard all terms where $\hat{e}_z$ appears. That is, to focus on the rows of the column vector $\vec{L}$ with   index  $n=1:\hat{e}_x\hat{e}_x\hat{e}_x, 2:\hat{e}_y\hat{e}_y\hat{e}_y, 8:(\hat{e}_x\hat{e}_y\hat{e}_y + \hat{e}_y\hat{e}_x\hat{e}_y +\hat{e}_y\hat{e}_y\hat{e}_x)~\text{and}~9:(\hat{e}_x\hat{e}_x\hat{e}_y + \hat{e}_x\hat{e}_y\hat{e}_x +\hat{e}_y\hat{e}_x\hat{e}_x)$. If the incident electric field is defined as $\vec{E}=|E|(\cos \theta \hat{e}_x + \sin \theta\hat{e}_y)$ after substitution in equation \ref{eq:MChi} the intensities of the THG are obtained:

\begin{eqnarray}
 I^{3\omega}_a \propto |E^{3\omega}_a|^2 \propto \left ( \chi_{11} \cos^3 \theta + \chi_{12}\sin^3 \theta + 3\chi_{18}\cos\theta\sin^2\theta +3\chi_{19}\cos^2\theta\sin\theta \right )^2, \\
 I^{3\omega}_b \propto |E^{3\omega}_b|^2 \propto \left ( \chi_{21} \cos^3 \theta + \chi_{22}\sin^3 \theta + 3\chi_{28}\cos\theta\sin^2\theta +3\chi_{29}\cos^2\theta\sin\theta \right )^2.
\label{Int}
\end{eqnarray}

\noindent The contracted tensor components are then used as fitting parameters to fit the experimental data in Fig. \ref{Fig:few_mid}. Note that, particularly in the zoomed regions (Figs. \ref{Fig:few_mid}d,e,f), the analytical curves fit the experimental data satisfactorily. 
The $\chi_{mn}$ components, normalized by $\chi_{11}$, obtained after the least square fitting process is carried out, are presented in Table \ref{tablechi}. 

It is clearly observed that the elements of the $\chi^{(3)}$ tensor change with the sample thickness, explaining the observed differences in the lobes shapes. In particular, note that, as mentioned, the shoulder-like structure appears in the curve associated with $I_b^{3\omega}$ (equation \ref{Int}), for which the tensor index $m=2$. Furthermore, the shoulder is obtained when the pump is practically aligned with the $x$ axis. Thus, we can, to a first approximation, neglect indices $n$ corresponding to two, or more, $y$ pump components, relative to indices corresponding to two, or more, $x$ pump components. This means that the most relevant $n$ indices are $n=1$ and $n=9$. Indeed, inspection of Table \ref{tablechi} shows that $mn$ components 21 and 29, in fact, present appreciable values (up to $\sim30\%$ and $\sim10\%$ of the value of $\chi_{11}$, respectively) and a monotonic variation as the thickness increases ($\chi_{21}$ drops and $\chi_{29}$ increases). Table \ref{tablechi} also includes a comparison with the results reported by Tripathi \textit{et al.}\cite{tripathi2021naturally}. Note that, in this case, the authors have only measured flakes down to a thickness of 25 nm and noticed no significant changes in the tensor components. The values provided are, therefore, thickness-averaged values. Indeed, they exhibit a fair agreement with the values we found, especially for the case of our two thicker flakes, for which case $\chi_{18}$, $\chi_{19}$, $\chi_{21}$, and $\chi_{28}$ show lower values than for the 10-nm-thick sample. Note that changes in the amplitude (and phase) of the $\chi^{(3)}$  tensor components ultimately trace back to changes in the spectral position of resonances and/or to the oscillator strength between different electronic states in franckeite. Therefore, the thickness-dependent changes reported here are like to be a consequence of changes in the electronic structure and, consequently can also impact the performance of photonic and optoelectronic devices (e.g. photodetectors and optical modulators \cite{molina2017franckeite, RayACS2017, LiNanophot2021}) based on franckeite. However, quantification of such an impact is beyond the scope of this paper.



\begin{table}[h]
\begin{center}
\begin{tabular}{|c|c|c|c|c|c|c|c|c|c|}
\hline
\textbf{Sample} & $\chi_{11}$ & $\chi_{12}$ & $\chi_{18}$ &$\chi_{19}$&$\chi_{21}$&$\chi_{22}$&$\chi_{28}$&$\chi_{29}$ \\ [0.5ex] 
\hline
10 nm       &  1             &     0                & 0.0719       & 0.0424        & 0.3454         &  0.8692       &  0.0748    & 0.0370 \\ [1ex] 
34 nm       &  1             &     0.0052       & 0.0476       & 0.0018        & 0.2819         &  0.7796       &  0.2139    & 0.0638 \\ [1ex] 
124 nm       &  1             &     0       & 0.0633       & 0        & 0         &   0.9491       &  0    & 0.1084\\ [1ex] 
Tripathi \textit{et al.}\cite{tripathi2021naturally}&  1             &     0.012	    & 0.043       & 0.015        & 0.017         &   0.815       &  0.009    & 0.037\\[0.5ex] 
\hline
\end{tabular}
\end{center}
\caption{Values of $\chi_{mn}$ normalized by $\chi_{11}$ obtained by the least-square fitting process, and comparison with Ref. \cite{tripathi2021naturally}}
\label{tablechi}
\end{table}

\begin{figure}[ht]
\centering
  \includegraphics[width=0.5\textwidth]{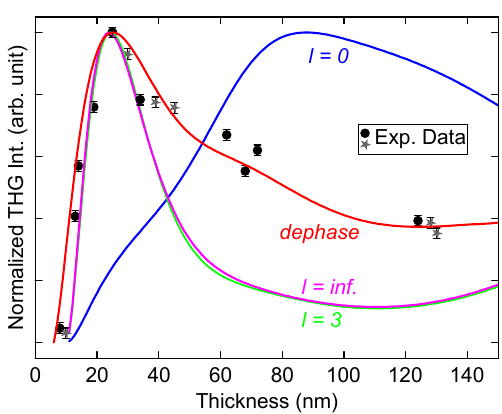}
  \caption{THG intensity plotted versus franckeite thickness (symbols). Thicknesses were extracted from AFM measurements in two staircase-like franckeite samples (black circles and gray star  symbols represent the different samples measured). The solid lines correspond to the best fit analytical curves obtained with various models: (purple) fully coherent model with infinite number of reflections; (blue) coherent model with no reflections (the intensity was multiplied by 7.97 to match the maximum experimental intensity); (green) coherent model with $l=3$ reflections; and (red) partially coherent model. All models are explained in the text. The error bars shown were calculated as the standard deviation of 5 measurements taken at the same flake thickness.}
  \label{Fig:Thickness}
\end{figure}

We now analyze the influence of the material thickness on the maximum THG intensity (obtained, as discussed, for a pump polarization along the $a$ axis). Fig. \ref{Fig:Thickness} plots the dependence of the THG intensity on the franckeite thickness for the incident linear polarization along the rippling direction. In order to maintain consistency between all the points collected, even those that were in different flakes, the measurements were collected using the careful calibration criteria described in Methods. Color symbols express experimental data with error bars, and analytical fittings (obtained as discussed in Methods) are plotted with solid lines. Compared to previously reported measurements on thicker samples, where a wide Gaussian-like shape peak is seen around $\sim100$ nm,\cite{tripathi2021naturally} in our samples, the THG dependence on thickness shows a more complex structure. Surprisingly, the peak (maximum THG value) appears for much thinner samples, being at around 25 nm, with a shoulder-like feature located at about 70 nm.

To understand the observed trends, we model the sample as a layered structure and use the transfer matrix method for harmonic generation outlined in the Methods section. 
At normal incidence and with the polarization of the pump parallel to the rippling axis, we can, with no loss of generality, treat the THG fields as a one-dimensional problem. 
First, it is clear from equation \ref{eq:THGE3} that the source field determines the THG fields in the air and the silica glass regions. With this in mind, it seems reasonable to assume that multiple pump reflections within the flake cannot be neglected. We initially consider an infinite number of pump reflections within the nonlinear slab; that is, using for $r^{\omega}$ in equation \ref{eq:EmasEmn} the Airy reflection formula for three media ($r^{\omega} = \frac{r^{\omega}_{12} + r^{\omega}_{23}e^{i2N_2(\omega)(\omega/c)z}}{1+r^{\omega}_{12}r^{\omega}_{23}e^{i2N_2(\omega)(\omega/c)z}}$). Second, it is equally clear that the process of calculating the THG fields can be reversed and the optical constants of the material can be determined from the measurement of the THG fields. Thus, reversing the process we obtain the refractive index values $N_2(\omega) = 4.1159 + i0.5$ and $N_2(3\omega) = 3.5 + i1.5$, which satisfactorily reproduces the THG thickness dependence for thicknesses smaller than 30 nm (purple line in Fig. \ref{Fig:Thickness}). Indeed, these refractive index values are very similar to those previously reported\cite{tripathi2021naturally,gant2017optical}. The peak position of THG versus thickness curve may be attributed to phase mismatching between the pump and the third-harmonic wave. Indeed, the coherence length ($L_c$) of nonlinear process is given by\cite{TLow} $L_c = \lambda/6(N_2(\omega) + N_2(3\omega)) = 33.920$ nm for backward propagating waves and $L_c = \lambda/6(N_2(\omega)-N_2(3\omega)) = 419.44$ nm for the forward propagating waves. As multiple reflections take place, forward and backward-generated waves become co-propagating, meaning that the shorter $L_c$ dominates. Indeed, the length for backward THG closely matches the THG peak position.

In an effort to improve the fit to the experimental data in the thicker sample range, we evaluate the effect of limiting the number of reflections inside the nonlinear layer that effectively contributes to THG. This is done by the recursion formula $r^{\omega} = r^{\omega}_{12}+t^{\omega}_{21}r^{\omega}_{23}t^{\omega}_{12}\phi^{\omega}_2\sum_{j=0}^{l}q^j$, where $q = r^{\omega}_{21}r^{\omega}_{23}\phi^{\omega}_2$ and $\phi^{\omega}_2=e^{-i2N_2(\omega/c)d}$. For $l=0$, i.e., completely disregarding multiple reflections of the pump field, it is impossible to fit the experimental data (blue curve in Fig. \ref{Fig:Thickness}). This model gives a maximum THG intensity at $\sim$85 nm. It is important to mention that the calculated intensities for this case were scaled (x 7.97) in an effort to better fit the data. For $l=3$ (green curve in Fig. \ref{Fig:Thickness}), the peak THG value increases and moves towards smaller thicknesses, rapidly approaching the reaching the value for $l=\infty$. For $l > 5$ the curve for infinite reflections is recovered. However, the shoulder feature that is observed in the experiments can never be reproduced.

Note that the entire nonlinear optical treatment provided here assumes the pump to be monochromatic, consisting of one single frequency $\omega$. In practice, being a femtosecond pulse source, the pump is composed of a band of frequencies around $\omega$, which significantly scales the complexity of the problem. For example, sum-frequency generation involving different pump frequencies occurs, which can further reduce the coherence of the process. To emulate these features without resorting to a more complex model, we empirically add dephasing factors that affect the pump, by multiplying its Fresnel reflection (transmission) coefficients\cite{Mitsas:95,Katsidis:02}. The modified coefficients are now rewritten as $r_{ij} = \alpha r_{ij}^0$ and $t_{ij} = \beta t_{ij}^0$, where the superscript $0$ indicates the Fresnel coefficient for the monochromatic case. We find that with $r^{\omega}_{12} = 0.5 \left(r^{\omega}_{12}\right)^0$, $t^{\omega}_{21} = 0.5\left( t^{\omega}_{21}\right)^0$ and $r^{\omega}_{23} = 0.9 \left(r^{\omega}_{23}\right)^0$ the calculated intensities satisfactorily fit the experimental data for thin and thick samples, as can be seen for the red curve in Fig. \ref{Fig:Thickness}. Finally, note that if dephasing increases (with the factors tending to zero), the curve for $l=0$ (i.e., single pass THG) is recovered. We notice that the single-pass model was, indeed, used to fit the data reported by Tripathi \textit{et al.}\cite{tripathi2021naturally}, with a good experimental agreement. We speculate that this is so because those authors used a pump consisting of significantly shorter pulses (90 fs), for which case dephasing increases, approaching the single-pass picture.


\subsection{Conclusion}

In summary, we report on an experimental and theoretical investigation of the nonlinear optical properties of layered franckeite. The third-order nonlinear optical response (third-harmonic generation) was found to be highly anisotropic, varying in a nontrivial way with the incident polarization. We demonstrated that THG measurements are able to quickly identify the franckeite rippling axis with high accuracy and with no need for additional measurements. Additionally, we found the THG to be strongly dependent on the material thickness, not only in terms of absolute achieved harmonic intensity, but also in terms of pump polarization dependence. We attribute the latter to changes in the third-order nonlinear susceptibility tensor. A maximum THG intensity was observed at a franckeite thickness of about 25 nm, which is mainly attributed to multiple pump reflections within the franckeite flake and to phase matching.  
We believe that our findings contribute to a more comprehensive understanding of anisotropic layered materials and can potentialize their use in future photonic and optoelectronic applications.

\begin{acknowledgement}
 All authors acknowledge the Coordenação de Aperfeiçoamento de Pessoal de Nível Superior (CAPES), Conselho Nacional de Desenvolvimento de Científico e Tecnológico (CNPq) and the São Paulo State Foundation (FAPESP; grant numbers: 2020/04374-6, 2018/25339-4, and 2015/11779-4), and the Brazilian Nanocarbon Institute of Science and Technology (INCT/Nanocarbono). A.R.C., A.S.M.V.O., D.A.B., and C.J.S.d.M. acknowledge the support from Fundo Mackenzie de Pesquisa e Inovação (MackPesquisa grant number 221017).

\end{acknowledgement}





\bibliography{achemso-demo}

\end{document}